\documentclass{symmetry08rop}

\begin{document}

\allowdisplaybreaks
\FirstPageHeading{BoykoPateraPopovych}

\ShortArticleName{Invariants of Lie Algebras via Moving Frames}

\ArticleName{Invariants of Lie Algebras via Moving Frames}

\Author{Vyacheslav BOYKO~$^\dag$, Jiri PATERA~$^\ddag$ and Roman POPOVYCH~$^{\dag\S}$}
\AuthorNameForHeading{V.~Boyko, J.~Patera and R.~Popovych}

\AuthorNameForContents{Boyko V., Patera J.\ and Popovych R.}
\ArticleNameForContents{Invariants of Lie Algebras via Moving Frames Approach}

\Address{$^\dag$~Institute of Mathematics of NAS of Ukraine,\\
\hphantom{$^\dag$}~3 Tereshchenkivs'ka Str., Kyiv, 01004 Ukraine}
\EmailD{boyko@imath.kiev.ua, rop@imath.kiev.ua}

\Address{$^\ddag$~Centre de Recherches Math\'ematiques,
Universit\'e de Montr\'eal,\\
\hphantom{$^\ddag$}~C.P. 6128 succursale Centre-ville, Montr\'eal (Qu\'ebec), H3C 3J7 Canada}
\EmailD{patera@CRM.UMontreal.CA}

\Address{$^\S$~Faculty of Mathematics, University of Vienna,\\ 
\hphantom{$^\S$}~Oskar-Morgenstern-Platz 1, 1090 Vienna, Austria}

\Abstract{A purely algebraic algorithm for computation of
invariants (generalized Casimir operators) of Lie algebras by means of moving frames is discussed.
Results on the application of the method to computation of invariants of low-dimensional Lie algebras and
series of solvable Lie algebras restricted only by a required structure of the nilradical are reviewed.}

\section{Introduction}

The invariants of Lie algebras are one of their defining
characteristics. They have numerous applications in different fields
of mathematics and physics, in which Lie algebras arise
(representation theory, integrability of Hamiltonian differential
equations, quantum numbers etc). In particular, the polynomial
invariants of a~Lie algebra exhaust its set of Casimir operators,
i.e., the center of its universal enveloping algebra. This is why
non-polynomial invariants are also called generalized Casimir
operators, and the usual Casimir operators are seen as `specific'
generalized Casimir operators. Since the structure of invariants
strongly depends on the structure of the algebra and the
classification of all (finite-dimensional) Lie algebras is an
inherently difficult problem (actually unsolvable\footnote{The problem of classification of 
Lie algebras is \emph{wild} since it includes, as a subproblem, the problem on reduction
of pairs of matrices to a canonical form~\cite{BPP:Kirillov}. 
For a detailed review on classification of Lie algebras we refer to \cite{BPP:Popovych&Boyko&Nesterenko&Lutfullin2003}.}), 
it seems to be impossible to elaborate a complete theory for
generalized Casimir operators in the general case. Moreover, if the
classification of a class of Lie algebras is known, then the
invariants of such algebras can be described exhaustively. These
problems have already been solved for the semi-simple and
low-dimensional Lie algebras, and also for the physically relevant
Lie algebras of fixed dimensions.

The standard method of construction of generalized Casimir operators consists of integration
of overdetermined systems of first-order linear partial differential equations.
It turns out to be rather cumbersome calculations,
once the dimension of Lie algebra is not one of the lowest few.
Alternative methods use matrix representations of Lie algebras.
They are not much easier and are valid for a~limited class of representations.

In our recent papers 
\cite{BPP:Boyko&Patera&Popovych2006,BPP:Boyko&Patera&Popovych2007a,BPP:Boyko&Patera&Popovych2007b,
BPP:Boyko&Patera&Popovych2007c,BPP:Boyko&Patera&Popovych2008} 
we have developed the purely algebraic algorithm for computation of invariants (generalized Casimir operators) of Lie algebras.
The suggested approach  is simpler and generally valid.
It extends to our problem the exploitation of
the Cartan's method of moving frames in Fels--Olver version \cite{BPP:Fels&Olver1998-1999}.
(For modern development of the moving frames method and more references see also~\cite{BPP:Olver2008,BPP:Olver&Pohjanpelto2008}.)

\section{Preliminaries}\label{BPP:SectionPreliminaries}

Consider a Lie algebra~$\mathfrak g$ of dimension $\dim \mathfrak g=n<\infty$
over the complex or real field $\mathbb F$ (either $\mathbb F=\mathbb C$ or $\mathbb F=\mathbb R$) and the corresponding connected Lie group~$G$.
Let~$\mathfrak g^*$ be the dual space of the vector space~$\mathfrak g$.
The map ${\rm Ad}^*\colon G\to {\rm GL}(\mathfrak g^*)$ defined for any $g\in G$ by the relation
\[
\langle{\rm Ad}^*_g x,u\rangle=\langle x,{\rm Ad}_{g^{-1}}u\rangle
\quad \mbox{for all $x\in \mathfrak g^*$ and $u\in \mathfrak g$}
\]
is called the {\it coadjoint representation} of the Lie group~$G$.
Here ${\rm Ad}\colon G\to {\rm GL}(\mathfrak g)$ is the usual adjoint
representation of~$G$ in~$\mathfrak g$, and the image~${\rm Ad}_G$
of~$G$ under~${\rm Ad}$ is the inner automorphism group ${\rm
Int}(\mathfrak g)$ of the Lie algebra~$\mathfrak g$. The image
of~$G$ under~${\rm Ad}^*$ is a subgroup of~${\rm GL}(\mathfrak g^*)$ and
is denoted by~${\rm Ad}^*_G$.

The maximal dimension of orbits of~${\rm Ad}^*_G$ is called 
the {\em rank of the coadjoint representation} of~$G$ (and~$\mathfrak g$) 
and denoted by ${\mathop{\rm rank}\nolimits} {\rm Ad}^*_G$.
It is a basis independent characteristic of the algebra~$\mathfrak g$.
Orbits of this dimension are called regular ones. 

A~function $F\in C^\infty(\Omega)$, where $\Omega$ is a domain in~$\mathfrak g^*$, 
is called a (global in $\Omega$) {\it invariant} of~${\rm Ad}^*_G$  if
$
F({\rm Ad}_g^* x)=F(x)\ \mbox{for all}\ g\in G \ \mbox{and}\ x\in\Omega \ \mbox{such that}\ {\rm Ad}^*_g x\in\Omega.
$
The set of invariants of ${\rm Ad}^*_G$ on $\Omega$ is denoted by ${\mathop{\rm Inv}\nolimits}({\rm Ad}^*_G)$
without an explicit indication of the domain~$\Omega$. 
Let below $\Omega$ is a neighborhood of a point from a regular orbit. 
It can always be chosen in such a way that the group~${\rm Ad}^*_G$ acts regularly on $\Omega$. 
Then the maximal number $N_\mathfrak g$ of functionally independent invariants in 
${\mathop{\rm Inv}\nolimits}({\rm Ad}^*_G)$ coincides with the codimension of the regular orbits of~${\rm Ad}^*_G$, 
i.e., it is given by the difference
$
N_\mathfrak g=\dim \mathfrak g-{\mathop{\rm rank}\nolimits}\, {\rm Ad}^*_G.
$

To calculate the invariants explicitly, one should fix a basis
$\mathcal E=(e_1,\ldots,e_n)$ of the algebra~$\mathfrak g$. 
It leads to fixing the dual basis $\mathcal E^*=(e_1^*,\ldots,e_n^*)$ in the dual space~$\mathfrak g^*$ 
and to the identification of ${\rm Int}(\mathfrak g)$ and ${\rm Ad}^*_G$ with the associated matrix groups. 
The basis elements $e_1,\ldots,e_n$ satisfy the commutation relations $[e_i,e_j]=c_{ij}^k e_k$,
where $c_{ij}^k$ are components of the tensor of structure constants of~$\mathfrak g$ in the basis~$\mathcal E$.
Here and in what follows the indices $i$, $j$ and $k$ run from~1 to~$n$
and the summation convention over repeated indices is used.
Let $x\to\check x=(x_1,\ldots,x_n)$ be the coordinates in~$\mathfrak g^*$ associated with~$\mathcal E^*$. 

It is well known that there exists a bijection between 
elements of the center of the universal enveloping algebra (i.e., \emph{Casimir operators}) of~$\mathfrak g$ 
and polynomial invariants of~$\mathfrak g$ (which can be assumed defined globally on~$\mathfrak g^*$).
See, e.g., \cite{BPP:Abellanas&MartinezAlonso1975}.
Such a bijection is established, e.g., by the symmetrization operator~$\mathop{\rm Sym}$ 
which acts on monomials by the formula
\[
\mathop{\rm Sym}\nolimits (e_{i_1}\cdots e_{i_r})=\dfrac1{r!}\sum_{\sigma\in {\rm S}_r}e_{i_{\sigma_1}}\cdots e_{i_{\sigma_r}},
\]
where $i_1, \ldots, i_r$ take values from 1 to $n$, $r\in \mathbb N$.
The symbol ${\rm S}_r$ denotes the symmetric group on $r$~letters. 
The symmetrization also can be correctly defined for rational invariants~\cite{BPP:Abellanas&MartinezAlonso1975}. 
If ${\rm Int}({\rm Ad}^*_G)$ has no a functional basis consisting of only rational invariants, 
the correctness of the symmetrization needs an additional investigation for each fixed algebra~$\mathfrak g$ 
since general results on this subject do not exist. 
After symmetrized, elements from ${\rm Int}({\rm Ad}^*_G)$ are naturally called invariants or 
\emph{generalized Casimir operators} of~$\mathfrak g$.
The set of invariants of $\mathfrak g$ is denoted by ${\mathop{\rm Inv}\nolimits}(\mathfrak g)$.

Functionally independent invariants $F^l(x_1,\ldots,x_n)$,
$l=1,\ldots,N_\mathfrak g$, forms a {\it functional basis}
({\it fundamental invariant}) of ${\mathop{\rm Inv}\nolimits}({\rm Ad}^*_G)$ since any
element from ${\mathop{\rm Inv}\nolimits}({\rm Ad}^*_G)$ can be (uniquely) represented as
a~function of these invariants.
Accordingly the set of $\mathop{\rm Sym}\nolimits F^l(e_1,\ldots,e_n)$,
\mbox{$l=1,\ldots,N_\mathfrak g$}, is called a basis of~${\mathop{\rm Inv}\nolimits}(\mathfrak g)$.

In framework of the infinitesimal approach
any invariant $F(x_1,\ldots,x_n)$ of~${\rm Ad}^*_G$ is a solution of
the linear system of first-order partial differential equations
\cite{BPP:Abellanas&MartinezAlonso1975,BPP:Beltrametti&Blasi1966, BPP:Patera&Sharp&Winternitz&Zassenhaus1976}
$
X_iF=0$, i.e., $c_{ij}^k x_kF_{x_j}=0,$
where $X_i=c_{ij}^k x_k\partial_{x_j}$ is the infinitesimal generator
of the local one-parameter group $\{{\rm Ad}^*_{\exp(\varepsilon e_i)}\}$
corresponding to $e_i$, where the parameter~$\varepsilon$ runs through a neighborhood of zero in~$\mathbb F$. 
The mapping $e_i\to X_i$ gives a representation of the Lie algebra~$\mathfrak g$.

\section{The algorithm}\label{BPP:SectionAlgorithm}


Let~$\mathcal{G}={\rm Ad}^*_G\times \mathfrak
g^*$ denote the trivial left principal ${\rm Ad}^*_G$-bundle
over~$\mathfrak g^*$. The right regularization~$\widehat R$ of the
coadjoint action of~$G$ on~$\mathfrak g^*$ is the diagonal action
of~${\rm Ad}^*_G$ on~$\mathcal{G}={\rm Ad}^*_G\times \mathfrak g^*$.
It is provided by the map
\[
\widehat R_g({\rm Ad}^*_h,x)=({\rm Ad}^*_h\cdot {\rm Ad}^*_{g^{-1}},{\rm Ad}^*_g x),\quad
g,h\in G,\quad x\in \mathfrak g^*.
\]
The action~$\smash{\widehat R}$ on the bundle~$\mathcal{G}={\rm Ad}^*_G\times
\mathfrak g^*$ is regular and free. We call $\smash{\widehat R}$ the
\emph{lifted coadjoint action} of~$G$. It projects back to the
coadjoint action on~$\mathfrak g^*$ via the ${\rm
Ad}^*_G$-equivariant projection~$\pi_{\mathfrak g^*}\colon
\mathcal{G}\to \mathfrak g^*$. Any \emph{lifted invariant} of~${\rm
Ad}^*_G$ is a~(locally defined) smooth function from~$\mathcal{G}$
to a~manifold, which is invariant with respect to the lifted
coadjoint action of~$G$. The function $\mathcal
I\colon\mathcal{G}\to \mathfrak g^*$ given by $\mathcal I=\mathcal
I({\rm Ad}^*_g,x)={\rm Ad}^*_g x$ is the \emph{fundamental lifted
invariant} of ${\rm Ad}^*_G$, i.e., $\mathcal I$ is a~lifted
invariant and any lifted invariant can be locally written as a
function of~$\mathcal I$ in a unique way. Using an arbitrary function~$F(x)$
on~$\mathfrak g^*$, we can produce the lifted
invariant~$F\circ\mathcal I$ of~${\rm Ad}^*_G$ by replacing $x$ with
$\mathcal I={\rm Ad}^*_g x$ in the expression for~$F$. Ordinary
invariants are particular cases of lifted invariants, where one
identifies any invariant formed as its composition with the standard
projection~$\pi_{\mathfrak g^*}$. Therefore, ordinary invariants are
particular functional combinations of lifted ones that happen to be
independent of the group parameters of~${\rm Ad}^*_G$.

The essence of the normalization procedure by Fels and Olver can be presented in the form of
on the following statement.

\begin{proposition}\label{BPP:PropositionOnNormalization}
Suppose that~$\mathcal I=(\mathcal I_1,\ldots,\mathcal I_n)$ is a fundamental lifted invariant,
for the lifted invariants $\mathcal I_{j_1}$, \ldots, $\mathcal I_{j_\rho}$ and some constants $c_1$, \ldots, $c_\rho$
the system $\mathcal I_{j_1}=c_1$, \ldots, $\mathcal I_{j_\rho}=c_\rho$ is solvable with respect to the parameters
$\theta_{k_1}$,~\ldots, $\theta_{k_\rho}$ and substitution of the
found values of $\theta_{k_1}$,~\ldots, $\theta_{k_\rho}$ into the
other lifted invariants results in $m=n-\rho$ expressions $\smash{\hat{\mathcal I}_l}$, $l=1,\dots,m$, depending only on $x$'s.
Then $\rho={\mathop{\rm rank}\nolimits}\, {\rm Ad}^*_G$, $m=N_\mathfrak g$
and $\hat{\mathcal I}_1$, \ldots, $\hat{\mathcal I}_m$ form a basis of ${\mathop{\rm Inv}\nolimits}({\rm Ad}^*_G)$.
\end{proposition}

The \emph{algebraic algorithm} for finding invariants of the Lie
algebra $\mathfrak g$ is briefly formulated in the following four
steps.

1.\ {\it Construction of the generic matrix $B(\theta)$ of~${\rm
Ad}^*_G$.} $B(\theta)$ is the matrix of an inner automorphism of the
Lie algebra~$\mathfrak g$ in the given basis $e_1$, \ldots, $e_n$,
$\theta=(\theta_1,\ldots,\theta_r)$ is a complete tuple of group
parameters (coordinates) of~$\mathop{\rm Int}(\mathfrak g)$, and
$r=\dim{\rm Ad}^*_G=\dim\mathop{\rm Int}(\mathfrak g)=n-\dim{\rm
Z}(\mathfrak g),$ where ${\rm Z}(\mathfrak g)$ is the center
of~$\mathfrak g$.

2.\ {\it Representation of the fundamental lifted invariant.} The
explicit form of the fundamental lifted invariant~$\mathcal
I=(\mathcal I_1,\ldots,\mathcal I_n)$ of ${\rm Ad}^*_G$ in the
chosen coordinates~$(\theta,\check x)$ in ${\rm Ad}^*_G\times
\mathfrak g^*$ is \mbox{$\mathcal I=\check x\cdot B(\theta)$}, i.e.,
\[
(\mathcal I_1,\ldots,\mathcal I_n)=(x_1,\ldots,x_n)\cdot B(\theta_1,\ldots,\theta_r).
\]

3.\ {\it Elimination of parameters by normalization}. We choose the
maximum possible number $\rho$ of lifted invariants $\mathcal
I_{j_1}$, \ldots, $\mathcal I_{j_\rho}$, constants $c_1$, \ldots,
$c_\rho$ and group parameters
$\theta_{k_1}$,~\ldots,~$\theta_{k_\rho}$ such that the equations
$\mathcal I_{j_1}=c_1$, \ldots, $\mathcal I_{j_\rho}=c_\rho$ are
solvable with respect to $\theta_{k_1}$,~\ldots,~$\theta_{k_\rho}$.
After substituting the found values of
$\theta_{k_1}$,~\ldots,~$\theta_{k_\rho}$ into the other lifted
invariants, we obtain $N_\mathfrak g=n-\rho$ expressions $F^l
(x_1,\ldots,x_n)$ without $\theta$'s.

4.\ {\it Symmetrization.} The functions $F^l(x_1,\ldots,x_n)$
necessarily form a basis of ${\mathop{\rm Inv}\nolimits}({\rm Ad}^*_G)$. They are
symmetrized to $\mathop{\rm Sym}\nolimits F^l(e_1,\ldots,e_n)$. It
is the desired basis of~${\mathop{\rm Inv}\nolimits}(\mathfrak g)$.

Our experience on the calculation of invariants of a wide range of
Lie algebras shows that the version of the algebraic method, which
is based on Proposition~\ref{BPP:PropositionOnNormalization}, is most
effective.
In particular, it provides finding the cardinality of the invariant
basis in the process of construction of the invariants.
The algorithm can in fact involve different kinds of coordinate in the inner automorphism groups
(the first canonical, the second canonical or special one) and
different techniques of elimination of parameters (empiric techniques,
with additional combining of lifted invariants,
using a floating system of normalization equations etc).

Let us underline that the search of invariants of a Lie algebra $\mathfrak{g}$,
which has been done by solving a linear system of first-order partial differential equations
under the conventional infinitesimal approach,
is replaced here by the construction of the matrix~$B(\theta)$ of inner automorphisms
and by excluding the parameters~$\theta$
from the fundamental lifted invariant $\mathcal{I}=\check x\cdot B(\theta)$ in some way.

\section{Illustrative example}\label{BPP:SectionIllustrativeExample}

The four-dimensional solvable Lie algebra~$\mathfrak{g}_{4.8}^b$ has the following nonzero commutation relations
\begin{gather*}
[e_2,e_3]=e_1,  \quad [e_1,e_4]=(1+b)e_1, \quad
[e_2,e_4]=e_2 , \quad [e_3,e_4]=be_3, \quad |b|\leq 1.
\end{gather*}
Its nilradical is three-dimensional and isomorphic to the Weil--Heisenberg algebra~$\mathfrak{g}_{3.1}$.
(Here we use the notations of low-dimensional Lie algebras according to Mubarakzyanov's classification~\cite{BPP:mubarakzyanov1963.1}.)

We construct a presentation of the inner automorphism matrix $B(\theta)$ of the Lie algebra~$\mathfrak g$, involving
second canonical coordinates on ${\rm Ad}_G$ as group parameters~$\theta$.
The matrices~$\hat{\rm ad}_{e_i}$, $i=1,\dots,4,$ of the adjoint representation
of the basis elements $e_1$, $e_2$, $e_3$ and $e_4$ respectively have the form
\begin{gather*}
\left(\begin{array}{cccc} 0& 0 & 0 & 1+b \\ 0&0&0&0\\0&0&0&0\\ 0&0&0&0 \end{array}\right)\!, \quad
\left(\begin{array}{cccc} 0& 0 & 1 & 0 \\ 0&0&0&1\\0&0&0&0\\ 0&0&0&0 \end{array}\right)\!, \\[1ex]
\left(\begin{array}{cccc} 0& -1 & 0 & 0 \\ 0&0&0&0\\0&0&0&b\\ 0&0&0&0 \end{array}\right)\!, \quad
\left(\begin{array}{cccc} -1-b& 0 & 0 & 0 \\ 0&-1&0&0\\0&0&-b&0\\ 0&0&0&0 \end{array}\right)\!.
\end{gather*}
The inner automorphisms of~$\mathfrak{g}_{4.8}^{b}$ are then described by the triangular matrix
\begin{gather*}
B(\theta)=\prod_{i=1}^3\exp(\theta_i\hat{\rm ad}_{e_i}) \cdot \exp(-\theta_4\hat{\rm ad}_{e_4})\\
\phantom{B(\theta)}{}
=\left(\begin{array}{@{}c@{\,\,\,}c@{\,\,\,}c@{\,\,\,}c@{}}
e^{(1+b)\theta_4} & -\theta_3 e^{\theta_4} & \theta_2 e^{b \theta_4}&
b\theta_2\theta_3+(1+b)\theta_1\\
0&e^{\theta_4}  & 0 &\theta_2 \\
0& 0 &e^{b \theta_4}   & b\theta_3 \\
0&0&0&1
\end{array}\right)\!.
\end{gather*}
Therefore, a functional basis of lifted invariants is formed by
\begin{gather*}
\mathcal{I}_1=e^{(1+b)\theta_4} x_1,\\
\mathcal{I}_2=e^{\theta_4}(-\theta_3  x_1+ x_2),\\
\mathcal{I}_3=e^{b\theta_4} (\theta_2 x_1 + x_3),\\
\mathcal{I}_4=(b\theta_2\theta_3+(1+b)\theta_1)x_1+\theta_2x_2+b\theta_3x_3+x_4.
\end{gather*}

Further the cases $b=-1$ and $b\not=-1$ should be considered separately.

There are no invariants in case $b\not=-1$ since in view of
Proposition~\ref{BPP:PropositionOnNormalization} the number of functionally independent invariants is equal to zero.
Indeed, the system $\mathcal I_1=1$, $\mathcal I_2=\mathcal I_3=\mathcal I_4=0$
is solvable with respect to the whole set of the parameters~$\theta$.

It is obvious  that in the case $b=-1$ the element $e_1$ generating the center~$Z(\mathfrak{g}_{4.8}^{-1})$ is an invariant.
(The corresponding lifted invariant $\mathcal{I}_1=x_1$ does not depend on the parameters~$\theta$.)
Another invariant is easily found via combining the lifted invariants:
$\mathcal{I}_1\mathcal{I}_4-\mathcal{I}_2\mathcal{I}_3=x_1x_4-x_2x_3$.
After the symmetrization procedure we obtain the following polynomial basis of the invariant set of this algebra
\[
e_1, \quad e_1e_4-\frac{e_2e_3+e_3e_2}{2}.
\]
The second basis invariant can be also constructed by the normalization technique.
We solve the equations $\mathcal I_2=\mathcal I_3=0$ with respect to the parameters $\theta_2$ and $\theta_3$ and
substitute the expressions for them into the lifted invariant $\mathcal{I}_4$.
The obtained expression $x_4-x_2x_3/x_1$ does not contain the parameters~$\theta$ and, therefore, is an invariant
of the coadjoint representation.
For the basis of invariants to be polynomial, we multiply this invariant by the invariant~$x_1$.
It is the technique that is applied below for the general case of the Lie algebras under consideration.

Note that in the above example the symmetrization procedure can be assumed trivial
since the symmetrized invariant $e_1e_4-\frac12(e_2e_3+e_3e_2)$ differs from the non-symmetrized version
$e_1e_4-e_2e_3$ (resp. $e_1e_4-e_3e_2$) on the invariant $\frac12e_1$ (resp. $-\frac12e_1$).
If we take the rational invariant $e_4-e_2e_3/e_1$ (resp. $e_4-e_3e_2/e_1$),
the symmetrization is equivalent to the addition of the constant~$\frac12$ (resp. $-\frac12$).

Invariants of~$\mathfrak{g}_{4.8}^{b}$ were first described in~\cite{BPP:Patera&Sharp&Winternitz&Zassenhaus1976} within the framework
of the infinitesimal approach.

\section{Review of obtained results}

Using the moving frames approach, we recalculated invariant bases and, in a number of cases, enhanced their representation for the following Lie algebras (in additional brackets we cite the papers where invariants bases of the same algebras were computed by the infinitesimal method):
\begin{itemize}\itemsep=-0.5ex
\item
the complex and real Lie algebras up to dimension 6~\cite{BPP:Boyko&Patera&Popovych2006} (\cite{BPP:Campoamor-Stursberg2005b,BPP:Ndogmo2000,BPP:Patera&Sharp&Winternitz&Zassenhaus1976});

\item
the complex and real Lie algebras with Abelian nilradicals of codimension one \cite{BPP:Boyko&Patera&Popovych2007a} (\cite{BPP:Snobl&Winternitz2005});

\item
the complex indecomposable solvable Lie algebras with the nilradicals isomorphic
to~${\mathfrak J}_0^n$, $n=3,4,\ldots\,$
(the nonzero commutation relations between the basis elements $e_1$, \dots, $e_n$ of ${\mathfrak J}_0^n$ are exhausted by
$[e_k,e_n]=e_{k-1}$, $k=2,\ldots,n-1$)~\cite{BPP:Boyko&Patera&Popovych2007a} (\cite{BPP:Ndogmo&Wintenitz1994b});

\item
the nilpotent Lie algebra $\mathfrak t_0(n)$ of $n\times n$ strictly upper triangular matri\-ces~\mbox{\cite{BPP:Boyko&Patera&Popovych2007a,BPP:Boyko&Patera&Popovych2007b}} (\cite{BPP:Tremblay&Winternitz2001});

\item
the solvable Lie algebra $\mathfrak t(n)$ of $n\times n$ upper triangular matrices
and the solvable Lie algebras $\mathfrak{st}(n)$ of $n\times n$ special upper triangular matrices \cite{BPP:Boyko&Patera&Popovych2007b,BPP:Boyko&Patera&Popovych2007c,BPP:Boyko&Patera&Popovych2008} (\cite{BPP:Tremblay&Winternitz2001});

\item
the solvable Lie algebras with nilradicals isomorphic to $\mathfrak t_0(n)$ and diagonal nilindependent elements
\cite{BPP:Boyko&Patera&Popovych2007b,BPP:Boyko&Patera&Popovych2007c,BPP:Boyko&Patera&Popovych2008} (\cite{BPP:Tremblay&Winternitz2001}).

\end{itemize}

Note that earlier only conjectures on invariants of two latter families of Lie algebras were known.
Moreover, for the last family the conjecture was formulated only for the particular case of a single nilindependent element.
Here we present the exhaustive statement on invariants of this series of Lie algebras, which was obtained in~\cite{BPP:Boyko&Patera&Popovych2008}.

Consider the solvable Lie algebra $\mathfrak t_{\gamma}(n)$
with the nilradical $\mathop{\rm NR}\nolimits(\mathfrak t_{\gamma}(n))$ isomorphic to $\mathfrak t_0(n)$ and
$s$ nilindependent element $f_p$, $p=1,\ldots,s$, which act on elements of the nilradical in the way
as the diagonal matrices $\Gamma_p={\mathop{\rm diag}\nolimits}(\gamma_{p1},\dots,\gamma_{pn})$ act on strictly triangular matrices.
The matrices $\Gamma_p$, $p=1,\ldots,s$, and the identity matrix are linear independent since otherwise
$\mathop{\rm NR}\nolimits(\mathfrak t_{\gamma}(n))\ne\mathfrak t_0(n)$.
The parameter matrix $\gamma=(\gamma_{pi})$ is defined up to nonsingular $s\times s$ matrix multiplier and
homogeneous shift in rows.
In  other words, the algebras $\mathfrak t_{\gamma}(n)$ and $\mathfrak t_{\gamma'}(n)$ are isomorphic if and only if
there exist $\lambda\in{\rm GL}(s,\mathbb F)$ and  $\mu\in\mathbb F^s$ such that
\begin{gather*}
\gamma'_{pi}=\lambda_{pp'}\gamma_{p'\!i}+\mu_p,\quad p=1,\dots,s,\ i=1,\dots,n,
\\[-.5ex]
\hspace*{-\mathindent}\mbox{or}\\[-.5ex]
\gamma'_{pi}=\lambda_{pp'}\gamma_{p'\!,n-i+1}+\mu_p,\quad p=1,\dots,s,\ i=1,\dots,n.
\end{gather*}
The parameter matrix $\gamma$ and $\gamma'$ are assumed equivalent.
Up to the equivalence the additional condition $\mathop{\rm tr}\Gamma_p=\sum_{i}\gamma_{pi}=0$ can be imposed on the algebra parameters.
Therefore, the algebra $\mathfrak t_{\gamma}(n)$ is naturally embedded into $\mathfrak{st}(n)$ as an ideal 
under identification of $\mathop{\rm NR}\nolimits(\mathfrak t_{\gamma}(n))$ with $\mathfrak t_0(n)$ and of $f_p$ with $\Gamma_p$.

We choose the concatenation of the canonical basis of $\mathop{\rm NR}\nolimits(\mathfrak t_{\gamma}(n))$ and the $s$-element set~$\{f_p,p=1,\ldots,s\}$
as the canonical basis of $\mathfrak t_{\gamma}(n)$.
In the basis of $\mathop{\rm NR}\nolimits(\mathfrak t_{\gamma}(n))$ we use `matrix' enumeration of basis elements
$e_{ij}$, $i<j$, with the `increasing' pair of indices similarly to
the canonical basis $(E^n_{ij},\,i<j)$ of the isomorphic matrix algebra $\mathfrak t_0(n)$.

Hereafter
$E^n_{ij}$ (for the fixed values $i$ and $j$) denotes the $n\times n$ matrix $(\delta_{ii'}\delta_{jj'})$
with $i'$ and $j'$ running the numbers of rows and column, respectively,
i.e., the $n\times n$ matrix with the unit on the cross of the $i$th row and the $j$th column and the zero otherwise.
The indices $i$, $j$, $k$ and $l$ run at most from~1 to~$n$.
Only additional constraints on the indices are indicated.
The subscript $p$ runs from~1 to~$s$, the subscript $q$ runs from~1 to~$s'$.
The summation convention over repeated indices $p$ and $q$ is used unless otherwise stated.
The number~$s$ is in the range $0,\dots,n-1$.
In the case $s=0$ we assume $\gamma=0$, and all terms with the subscript~$p$ should be omitted from consideration.
The value $s'$ ($s'<s$) is defined below.

Thus, the basis elements $e_{ij}\sim E^n_{ij}$, $i<j$, and $f_p\sim\sum_i \gamma_{pi}E^n_{ii}$ satisfy the commutation relations
$
[e_{ij},e_{i'\!j'\!}]=\delta_{i'\!j}e_{ij'\!}-\delta_{ij'}e_{i'\!j}, \
[f_p,e_{ij}]=(\gamma_{pi}-\gamma_{pj})e_{ij},
$
where $\delta_{ij}$ is the Kronecker delta.

The Lie algebra $\mathfrak t_{\gamma}(n)$ can be considered as the Lie algebra of the Lie subgroup
$
T_\gamma(n)=\{B\in T(n)\mid \exists \, \varepsilon_p\in\mathbb F\colon b_{ii}=e^{\gamma_{pi}\varepsilon_p}\}
$
of the Lie group $T(n)$ of non-singular upper triangular $n\times n$ matrices.

Below $A^{i_1,i_2}_{j_1,j_2}$, where $i_1\leqslant i_2$, $j_1\leqslant j_2$,
denotes the submatrix $(a_{ij})^{i=i_1,\ldots,i_2}_{j=j_1,\ldots,j_2}$ of a matrix $A=(a_{ij})$.
The conjugate value of $k$ with respect to $n$ is denoted by $\varkappa$, i.e., $\varkappa=n-k+1$.
The standard notation $|A|=\det A$ is used.

\begin{proposition}\label{BPP:PropositionOnReducedFormOfParameterMatrix}
Up to the equivalence relation on algebra parameters, the following conditions can be assumed satisfied
for some $s'\in\{0,\dots,\min(s,[n/2])\}$ and
$k_q,$ $q=1,\dots,s'$, $1\leqslant k_1<k_2<\dots<k_{s'\!}\leqslant[n/2]$:
\begin{gather*}
\gamma_{qk}=\gamma_{q\varkappa},\ k<k_q,\quad
\gamma_{q\varkappa_q}-\gamma_{qk_q}=1,\quad
\gamma_{pk_q}=\gamma_{p\varkappa_q},\ p\ne q,\quad q=1,\dots,s',
\\
\gamma_{pk}=\gamma_{p\varkappa},\ p>s',\ k=1, \ldots, [n/2],
\end{gather*}
where $\varkappa:=n-k+1$, $\varkappa_q:=n-k_q+1$.
\end{proposition}

We will say that the parameter matrix~$\gamma$ is of a \emph{reduced form}
if it satisfies the conditions of Proposition~\ref{BPP:PropositionOnReducedFormOfParameterMatrix}.

\begin{theorem}\label{TheoremOnBasisOfInvsOfDiagSolvAlgsWithTriangularNilradical}
Let the parameter matrix~$\gamma$ be of a reduced form.
A basis of~${\rm Inv}(\mathfrak t_\gamma(n))$ is formed by the expressions
\begin{gather*}
|\mathcal E^{1,k}_{\varkappa,n}|\prod_{q=1}^{s'\!}|\mathcal E^{1,k_q}_{\varkappa_q,n}|^{\beta_{qk}}, \quad
k\in\{1, \dots, [n/2]\}\setminus\{k_1 , \dots, k_{s'\!}\},
\\ \arraycolsep=.5ex
f_p+\sum_{k=1}^{\left[\frac n2\right]} \frac{(-1)^{k+1}}{|\mathcal E^{1,k}_{\varkappa,n}|}
(\gamma_{pk}-\gamma_{p,k+1}) \sum_{k<i<\varkappa}
\left|\begin{array}{lc} \mathcal E^{1,k}_{i,i} & \mathcal E^{1,k}_{\varkappa,n} \\[1ex]
0 & \mathcal E^{i,i}_{\varkappa,n} \end{array}\!\right|,
\quad p=s'+1, \ldots, s,
\end{gather*}
where 
$\varkappa:=n-k+1$, $\varkappa_q:=n-k_q+1$;
$\mathcal E^{i_1,i_2}_{j_1,j_2}$, $i_1\leqslant i_2$, $j_1\leqslant j_2$, denotes the matrix $(e_{ij})^{i=i_1,\ldots,i_2}_{j=j_1,\ldots,j_2}$;
$\beta_{qk}=-\Delta_{qk}/\Delta$, $\Delta=\det(\alpha_{q'\!k_{q''}})_{q'\!,q''\!=1,\dots,s'}=(-1)^{s'}$,
$\Delta_{qk}$ is the determinant obtained from $\Delta$
with change of the column $(\alpha_{q'\!k_q})_{q'\!=1,\dots,s'}$ by
the column $(\alpha_{q'\!k})_{q'\!=1,\dots,s'}$,
\[
\alpha_{qk}:=-\sum_{k'\!=1}^k (\gamma_{q,n-k'+1}-\gamma_{qk'\!})=-\sum_{k'\!=k_q}^k (\gamma_{q,n-k'+1}-\gamma_{qk'\!}).
\]
\end{theorem}

We use the short `non-symmetrized' form for basis invariants, where it is uniformly assumed
that in all monomials elements of $\mathcal E^{1,k}_{i,i}$ is placed before (or after)
elements of $\smash{\mathcal E^{i,i}_{\varkappa,n}}$.

\section{Conclusion}

The main advantage of the proposed method is in that it is
purely algebraic. Unlike the conventional infinitesimal method, it eliminates
the need to solve systems of partial differential equations, replaced
in our approach by the construction of the matrix~$B(\theta)$ of inner automorphisms
and by excluding the parameters~$\theta$
from the fundamental lifted invariant $\mathcal{I}=\check x\cdot B(\theta)$ in some way.

The efficient exploitation of the method imposes certain constraints on the choice of bases of the Lie algebras.
See, e.g., Proposition~\ref{BPP:PropositionOnReducedFormOfParameterMatrix} and Theorem~\ref{TheoremOnBasisOfInvsOfDiagSolvAlgsWithTriangularNilradical}.
That then automatically yields simpler expressions for the invariants.
In some cases the simplification is considerable.

\looseness=-1
Possibilities on the usage of the approach and directions for further investigation were outlined in our previous papers
\cite{BPP:Boyko&Patera&Popovych2006,BPP:Boyko&Patera&Popovych2007a,BPP:Boyko&Patera&Popovych2007b,BPP:Boyko&Patera&Popovych2007c,BPP:Boyko&Patera&Popovych2008}.
Recently advantages of the moving frames approach for computation of generalized Casimir operators were demonstrated in~\cite{BPP:Snobl&Winternitz2009}
with a new series of solvable Lie algebras.
The problem on optimal ways of applications of this approach to unsolvable Lie algebras is still open.

\medskip

{\bf Acknowledgments.}
We thank the members of the local organizing committee for the nice conference and for their hospitality.
J.P. work was supported by the Natural Science and Engineering
Research Council of Canada.
The research of R.P. was supported  by the Austrian Science Fund (FWF),  project P20632.

\end{document}